\documentclass[twocolumn]{aastex701}

\newcommand{\DM}{\mathrm{DM}}
\newcommand{\Cl}{C_\ell}
\newcommand{\yDM}{C_\ell^{y \times \DM}}

\usepackage{multirow}
\usepackage{amsmath} 
\usepackage{booktabs}

\begin{document}

\title{Beyond Feedback: Disentangling Baryonic Effects with tSZ$\times$FRB Cross-Correlations}

\author{Isabel Medlock}
\affiliation{Department of Astronomy, Yale University, New Haven, CT 06520, USA}
\email{isabel.medlock@yale.edu}

\author{Daisuke Nagai}
\affiliation{Department of Physics, Yale University, New Haven, CT 06520, USA}
\affiliation{Department of Astronomy, Yale University, New Haven, CT 06520, USA}
\email{daisuke.nagai@yale.edu}

\begin{abstract}

Standard observational probes lack the sensitivity to characterize the interplay between feedback-driven gas ejection and non-thermal pressure support in groups and clusters, and current models lack the framework to disentangle it. Recent first detections of the tSZ$\times$FRB cross-power spectrum \citep{Takahashi2025, Sharma_2026} have been interpreted as constraining AGN feedback and $\sigma_8$, but the baryon models used do not independently parameterize non-thermal pressure support, leaving feedback efficiency and gas thermodynamics degenerate. We demonstrate that this cross-correlation provides sensitivity to the thermal and non-thermal structure of cluster gas that neither observable alone can achieve, because the tSZ effect traces electron \textit{pressure} while FRBs trace solely electron \textit{density}. Using the Baryon Pasting (BP) framework, which separately parameterizes feedback efficiency $\epsilon_f$ and non-thermal pressure amplitude $A_{\rm nt}$, we show that their joint constraint breaks this degeneracy. In the noise-free limit, adding the cross-power spectrum reduces $r_{\rm cond}$ from $0.96$ to $0.08$ (a 12-fold reduction) and improves the joint figure of merit by a factor of 19, with the constraining power concentrated at cluster interior scales ($\ell \gtrsim 3000$). Recent detections are consistent with our fiducial model and disfavor weak feedback ($\epsilon_f^{95\%} \gtrsim 2.40\text{--}3.67 \times 10^{-6}$ across datasets). With $5\times10^4$ FRBs from DSA-2000 combined with SO, we forecast $3.8\%$ fractional precision on $A_{\rm nt}$, improving to $2.3\%$ with CMB-HD. These constraints directly inform the hydrostatic mass bias, with immediate implications for cluster-based cosmological inference from eROSITA, SO, and CMB-HD.

\end{abstract}

\section{Introduction}\label{sec:intro}

With the advent of increasingly sensitive multi-wavelength observations, including the thermal and kinetic Sunyaev-Zel'dovich (tSZ and kSZ) effects and fast radio bursts (FRBs), we now know that the majority of baryons reside in a diffuse, warm-hot phase permeating halos and the intergalactic medium (IGM) \citep[e.g.][]{Cen1999, deGraaff_2019, Connor2024}. The challenge has now shifted from merely detecting this gas to precisely characterizing both its three-dimensional distribution and thermodynamic state. This is essential for both enabling precision cosmology, where the uncertainty in baryonic physics is now the leading systematic for weak lensing at non-linear scales \citep{vanDaalen_2011, Chisari_2019, vanDaalen_2020, Amon_2022}, and understanding galaxy formation and evolution.

Feedback from stars and active galactic nuclei (AGN) heats and expels halo gas, shaping the density and temperature profiles of the intragroup and cluster media (IGrM and ICM). Yet these hot halos also host significant non-thermal pressure support from turbulence, bulk motions, magnetic fields, and cosmic rays \citep{Lau_2009, Angelinelli_2020, Green_2020}. Simulations indicate that non-thermal pressure is significant near the outskirts of clusters, which are more strongly affected by recent gas accretion \citep{Lau_2009, Nelson2014}, although the radial non-thermal pressure profile remains poorly constrained observationally \citep[e.g.][]{Eckert_2022}. Observables such as the tSZ effect \citep{Sunyaev1972}, sensitive to the thermal electron pressure of hot gas in galaxy groups and clusters, struggle to disentangle these effects: the auto-power spectrum can be suppressed both by stronger feedback and more non-thermal pressure support \citep{Shaw2010, Battaglia2012, Nelson2014}. The non-thermal pressure contribution also causes the hydrostatic mass bias, in which cluster masses inferred assuming purely thermal pressure support are systematically underestimated \citep[e.g.][for a review]{Nagai_2007a, Shi_2016, Pratt_2019}. Multi-wavelength cross-correlations offer a powerful route to jointly constraining cosmology and astrophysics, as different combinations with different sensitivities can break degeneracies among astrophysical parameters and between astrophysical and cosmological parameters that would otherwise remain entangled for a single observable \citep[e.g.][]{Schaan_2021, Amodeo_2021, Schneider_2022, Troster_2022, Bigwood_2024}. 

FRBs, milliseconds long extragalactic radio transients, provide a powerful new tool to confront this challenge \citep{Lorimer2007, Thornton2013, FRBs_2022}. Their dispersion measure (DM) traces the integrated column of free electrons along the line of sight and encodes the baryon distribution of the Universe. FRBs are sensitive to larger scales than kSZ and do not require disentangling density from velocity or temperature \citep{Reischke2026, Sharma_2026}. Beyond tracing the cosmological baryon density through the mean DM--redshift relation and resolving the missing baryon problem \citep{Ioka2003, Inoue2004, Macquart2020, Connor2024}, the fluctuations or scatter around this mean (and their cross correlation with other large-scale structure tracers) carry information about the effects of feedback on gas within and around halos \citep[e.g.][]{McQuinn2014, Medlock_2021, Nicola_2022, Medlock_2024, Medlock_2025b, Reischke_Hagstotz_2026, Sharma_2026}.

Leveraging the growing census of localized FRBs and high-resolution Compton y-maps, the tSZ$\times$FRB cross-correlation was first measured with the DMs of 133 localized FRBs and $y$-maps from \textit{Planck} at $4\sigma$ and the Atacama Cosmology Telescope at $1.5\sigma$ \citep{Takahashi2025}, and independently at $3.8\sigma$ with $3,455$ unlocalized CHIME FRBs and \textit{Planck} NILC $y$-map \citep{Sharma_2026, CHIME2_2026}. Interpreting these detections to extract physical constraints, however, requires frameworks that can separately account for thermal feedback and non-thermal pressure, which most existing models lack.

The frameworks used to interpret these recent measurements, HMcode \citep{Mead_2021} and BCEmu \citep{BCEmu_2019, BCEmu_2021}, parameterize feedback or the resulting gas distribution but lack treatment or variation of non-thermal pressure support, respectively. Initial interpretations of these measurements aim to constrain $\sigma_8$ (the amplitude of matter density fluctuations) and AGN feedback, but without explicit parameterization, this leads to non-thermal pressure being absorbed into these quantities. Because the tSZ effect is sensitive to both electron density and temperature, particularly in cluster outskirts where non-thermal pressure peaks, while FRBs trace only the electron density, we propose that their cross-power spectrum can be used to isolate temperature and break the feedback and non-thermal pressure degeneracy, similarly to how the cross-correlation between FRBs and the kSZ effect has been used to break the optical depth degeneracy \citep{Madhavacheril2019}. 

In this paper, we investigate whether the tSZ$\times$FRB power spectrum can break the degeneracy between baryonic feedback and non-thermal pressure support. We use the Baryon Pasting (BP) model \citep{Lau2025}, which paints physically motivated gas profiles onto dark matter halos \citep{Ostriker2005} at a fraction of the computational cost of hydrodynamical simulations. BP parameterizes thermal feedback ($\epsilon_f$) and total non-thermal pressure support ($A_{\rm nt}$) separately, with interpretable prescriptions, along with cosmological parameters. This separation allows us to study the response of the tSZ$\times$FRB cross power spectrum to each parameter and physical process individually, and assess how the cross-correlation distinguishes between feedback and non-thermal pressure. We first assess the maximal constraints possible on feedback and non-thermal pressure in the idealized, noise-free regime and then compare with current detections. Finally, we perform a realistic forecast for the coming decade, particularly timely given the $\sim 10^4$ localized FRBs per year expected from DSA-2000, CHORD, and the CHIME outriggers, and the high-sensitivity $y$-maps forthcoming from SO and, ultimately, CMB-HD.

This paper is structured as follows. We describe the Baryon Pasting model, the tSZ and FRB observables, and the halo-model formalism in Section~\ref{sec:methods}. Section~\ref{sec:results} presents the degeneracy-breaking analysis, the comparison with current detections, and the future survey forecasts. We discuss the physical interpretation in Section~\ref{sec:disc_int}, the limitations and systematics in Section~\ref{sec:disc_lim}, and the directions for future work in Section~\ref{sec:disc_fut}. Section~\ref{sec:conc} summarizes our findings.

\section{Methods}
\label{sec:methods}

\subsection{The Baryon Pasting Gas Model}\label{sec:bp}

The Baryon Pasting (BP) model paints gas density and pressure profiles onto dark matter halos, capturing the essential thermodynamics of the halo gas. The total gas pressure is assumed to be in hydrostatic equilibrium (HSE) with the Navarro-Frenk-White (NFW) \citep{Navarro1996} gravitational potential under a polytropic equation of state, with a steeper effective polytrope in the cool, dense core following \citet{Flender_2017}. We refer the reader to \citet{Lau2025} for the complete model description and its calibration against hydrodynamical simulations and X-ray observations. All profiles and power spectra in this work are computed with \texttt{diffgas}, a JAX-based implementation of the BP model in which every evaluation is JIT-compiled and fully differentiable.

The BP framework has developed over two decades, including the analytic polytropic gas model in hydrostatic equilibrium within an NFW potential \citep{Ostriker2005, Bode2009}, non-thermal pressure support \citep{Shaw2010}, a cool-core model calibrated against \textit{Chandra} X-ray observations of SZ-selected clusters \citep{Flender_2017}, and an extension to N-body particles to capture the cosmic web \citep{Osato2023}. \citet{Lau2025} unified these ingredients into a self-consistent model, validated against hydrodynamical simulations and X-ray data. The BP model complements the hydrodynamical simulation approach, which provides simulation-based priors, but is necessarily dependent on the specific subgrid physics \citep[e.g.][]{somerville_physical_2015}. The BP approach of parameterizing the energy budget is also complementary to the baryonification model and other commonly used simulation-calibrated halo-model codes that fit profiles for gas displacement \citep[e.g.][]{Mead_2021, BCEmu_2019, Schneider_2025}.

\subsubsection{Halo Profiles in the BP Model}

The normalization of the density and pressure profiles is set by numerically solving the energy and momentum conservation equations for the gas, with the final gas energy
\begin{equation}
E_{g,f} = E_{g,i} + \epsilon_\mathrm{DM}|E_\mathrm{DM}|
        + \epsilon_f M_\star c^2 + \Delta E_p,
\label{eq:energy}
\end{equation}
where $E_{g,i}$ and $E_{g,f}$ are the initial and final gas energies, $\Delta E_p$ is the work done by gas expansion, and $\epsilon_\mathrm{DM}|E_\mathrm{DM}|$ is the energy transferred by dynamical friction during mergers (fixed to zero in this work, following \citet{Lau2025}, as it is degenerate with $\epsilon_f$). The feedback efficiency $\epsilon_f$, the central feedback parameter of this paper, controls the energy injected by supernovae and active galactic nuclei per unit stellar mass formed, heating and ejecting gas from the halo. The stellar mass follows
\begin{equation}
M_\star(M_{500c}) = f_\star
\left( \frac{M_{500c}}{3\times10^{14}\,M_\odot} \right)^{-S_\star},
\label{eq:fstar}
\end{equation}
with $f_\star$ and $S_\star$ the normalization and mass slope of the stellar fraction.

The total pressure is decomposed into thermal and non-thermal components, $P_\mathrm{tot} = P_\mathrm{th} + P_\mathrm{nt}$, with the non-thermal fraction parameterized following \citet{Nelson2014},
\begin{equation}
f_\mathrm{nt}(r) \equiv \frac{P_\mathrm{nt}}{P_\mathrm{tot}}
= 1 - A_\mathrm{nt}\left[1 + \exp\!\left\{-\left(
\frac{r}{B_\mathrm{nt}\,R_{200m}}\right)^{\gamma_\mathrm{nt}}\right\}\right],
\label{eq:fnt}
\end{equation}
where $A_\mathrm{nt}$, $B_\mathrm{nt}$, and $\gamma_\mathrm{nt}$ set the amplitude, transition radius, and steepness of the total non-thermal pressure fraction profile. The \citet{Nelson2014} parameterization builds on the \citet{Shaw2010} model by introducing an exponential form and $R_{200m}$ scaling, which is motivated by the near-universality observed in hydrodynamic simulations. $A_\mathrm{nt}$ is the saturation value for the amplitude. The partition is applied after solving hydrostatically for the density profile, so the non-thermal pressure model has no effect on the density profiles. The non-thermal pressure is explicitly parameterized independently of feedback, which we discuss in Section~\ref{sec:disc_int}.

At large radii $f_\mathrm{nt} \to 1 - A_\mathrm{nt}$, so $A_\mathrm{nt}$ is a \emph{thermal} amplitude \citep{Lau2025}. Higher values signify more thermal support and a stronger tSZ signal at fixed total pressure. Note the opposite sign convention from the widely used $\alpha_0$ of \citet{Shaw2010}, where $A_\mathrm{nt} \approx 1 - \alpha_0$ at large radii. We treat $A_\mathrm{nt}$ as mass- and redshift-independent, following the near-universal form of \citet{Nelson2014}, so that any residual mass dependence is absorbed into the fitted amplitude, since the signal is dominated by a narrow halo mass range near $M \sim 10^{14}\,M_\odot$ at $z < 1$.

Cool cores are modeled following \citet{Flender_2017}, and the parameters are held fixed at their fiducial values. The radial extent of the halo gas is set by the boundary parameter $P_\mathrm{bound}$, in units of the splashback radius \citep{Aung2021}.

Table~\ref{tab:bp_params} lists all BP parameters, their physical meanings, and the fiducial values adopted from the \citet{Lau2025} calibration. The subset $\{\epsilon_f, f_\star, S_\star, A_\mathrm{nt}, P_\mathrm{bound}\}$ is varied in the inference of Section~\ref{sec:mcmc}; the remaining BP parameters are held fixed.

\begin{table*} 
\centering
\begin{tabular}{llccc}
\hline\hline
Parameter & Physical meaning & Equation & Fiducial Value & Prior\\
\hline
$\epsilon_f$ & Feedback efficiency from SNe and AGN & Eq. (1) & $3.97 \times 10^{-6}$ & [0.001, 50]$\times 10^{-6}$ \\
$\epsilon_{\rm DM}$ & Dark matter energy transfer to gas & Eq. (1) & $0.0$ & -- \\
$f_\star$ & Amplitude of the stellar mass fraction & Eq. (2) & $0.026$ & 0.026$\pm$0.003 \\
$S_\star$ & Mass slope of stellar mass fraction & Eq. (2) & $0.12$ & 0.12$\pm$0.10\\
$x_{\rm break}$ & Cluster core radius in $R_{500c}$ & -- & $0.195$ & -- \\
$\Gamma$ & Polytropic index outside the cluster core & -- & $1.2$ & -- \\
$\Gamma_0$ & Polytropic index within cluster core & -- & $0.1024$ & -- \\
$\beta_g$ & Redshift evolution of polytropic index within cluster core & -- & $1.72$ & -- \\
$A_{\rm nt}$ & Amplitude of non-thermal pressure fraction profile & Eq. (3) & $0.452$ & [0.01, 0.99]\\
$B_{\rm nt}$ & Transition scale of non-thermal pressure fraction in $R_{200m}$ & Eq. (3) & $0.841$ & -- \\
$\gamma_{\rm nt}$ & Logarithmic slope of non-thermal pressure fraction profile & Eq. (3) & $1.628$ & -- \\
$P_{\rm bound}$ & Boundary of the gas in halo in dark matter splashback radius & -- & $1.89$ & 1.89$\pm$0.189\\
\hline
\end{tabular}
\caption{BP model parameters, their physical meanings, the defining equation in which each appears, and their fiducial values. Fiducial values are adopted from the BP model calibration of \citet{Lau2025} (see also \citealt{Shaw2010, Nelson2014, Flender_2017}), with the gas boundary $P_{\rm bound}$ set to the splashback-radius value of \citet{Aung2021}. Radial parameters are quoted in units of $R_{500c}$, $R_{200m}$, or the splashback radius $R_{\rm sp}$ as indicated.}
\label{tab:bp_params}
\end{table*}

\subsection{Fast Radio Bursts and Dispersion Measure}

The key FRB observable is the dispersion measure ($\mathrm{DM}$), defined as the integrated electron density along the line of sight from source to observer, 
\begin{equation} \label{eq:dmdef}
\mathrm{DM} = \int_{0}^{d} \frac{n_{e}(l)}{1+z}\,dl,
\end{equation}
where $d$ is the proper distance, $n_e$ is the free electron number density, $z$ is the redshift, and $l$ is the proper path length. The Milky Way, the FRB host galaxy, the IGM, and the hot halo gas of intervening halos all contribute, so that
\begin{equation} \label{eq:dmbreakdown}
\mathrm{DM}_{\rm obs} = \mathrm{DM}_{\rm MW} + \mathrm{DM}_{\rm IGM} + \mathrm{DM}_{\rm halo} 
+ \frac{\mathrm{DM}_{\rm Host}}{1+z}.
\end{equation}
The relationship between redshift and extragalactic $\mathrm{DM}_{\rm cosmic} = \mathrm{DM}_{\rm IGM} + \mathrm{DM}_{\rm halo}$ has been well established theoretically and confirmed through observations and simulations \citep[e.g.][]{Macquart2020, James_2022}.

In \texttt{diffgas}, the electron density follows from the gas density as $n_e(r) = n_{\rm gas}(r)\,\mu/\mu_e$ (with $\mu = 0.588$ and $\mu_e = 1.136$ for a primordial hydrogen fraction $X_H = 0.76$) and the projected profiles are obtained by line-of-sight integration to $5\,R_{\rm 200m}$. 

\subsection{The thermal Sunyaev-Zel'dovich Effect and the Compton y Parameter}

The thermal Sunyaev--Zel'dovich (tSZ) effect arises from inverse Compton scattering of CMB photons off hot electrons in the intracluster medium (ICM) and circumgalactic medium (CGM) \citep{Sunyaev1972}.
The Compton-$y$ parameter along a line of sight in direction $\hat{n}$ is
\begin{equation}
  y(\hat{n}) = \frac{\sigma_T}{m_e c^2}
  \int P_\mathrm{th}\!\left(\chi\hat{n}, z\right) d\chi,
  \label{eq:y}
\end{equation}
where $\sigma_T$ is the Thomson cross-section, $m_e$ is the electron mass, $c$ is the speed of light, $P_\mathrm{th}$ is the thermal electron pressure, and the integral is along the comoving distance $\chi$. Because $y$ is proportional to the pressure, it is sensitive to both gas temperature and density.

In \texttt{diffgas}, the thermal pressure profile is $P_{\rm th}(r) = P_{\rm tot}(r)\,[1 -f_{\rm nt}(r)]$, where $f_{\rm nt}(r)$ is the non-thermal pressure fraction following the \citet{Nelson2014} model. The spherical Compton-$y$ profile is then
\begin{equation}
    y_{\rm sph}(r) = \frac{\sigma_T}{m_e c^2}\,p_e\,P_{\rm th}(r),
\end{equation}
where $p_e = (2X_H + 2)/(5X_H + 3)$ converts the gas pressure to the electron pressure.

\subsection{Halo Model Cross-Power Spectrum}

We compute the angular cross-power spectrum between the tSZ Compton-$y$ field and the FRB DM field using the halo model framework \citep{Cooray2002}. In this approach, it is assumed that all matter resides in virialized halos with a well-defined mass function $dn/dM$ and bias $b(M, z)$.

The angular cross-power spectrum at multipole $\ell$ is
\begin{equation}
  \yDM = \Cl^{y \times \DM, \,\mathrm{1h}} + \Cl^{y \times \DM, \,\mathrm{2h}},
  \label{eq:Cl}
\end{equation}
where the one-halo and two-halo terms are
\begin{align}
  \Cl^{\mathrm{1h}} &=
  \int_0^{z_\mathrm{max}} \frac{dV}{dz}\,dz
  \int dM\,\frac{dn}{dM}\,
  \tilde{y}_\ell(M,z)\,\widetilde{\rm DM}_\ell(M,z),
  \label{eq:Cl_1h}
\end{align}
\begin{equation}
\begin{aligned}
\Cl^{\mathrm{2h}} &=
  \int_0^{z_\mathrm{max}} \frac{dV}{dz}\,P_\mathrm{lin}(k_\ell, z)\,dz \\
  &\quad \times
  \left[\int dM\,\frac{dn}{dM}\,b(M,z)\,\tilde{y}_\ell\right] \\
  &\quad \times
  \left[\int dM\,\frac{dn}{dM}\,b(M,z)\,\widetilde{\DM}_\ell\right].
\end{aligned}
\label{eq:Cl_2h}
\end{equation}
Here $dV/dz$ is the comoving volume element per steradian per unit redshift, $P_\mathrm{lin}(k,z)$ is the linear matter power spectrum evaluated at $k = (\ell + 1/2)/\chi(z)$ under the Limber approximation \citep{Limber1953}, and $b(M,z)$ is the linear halo bias \citep{Tinker2010}. The mass function $dn/dM$ is taken from \citet{Tinker2008}.

We present results as $D_\ell \equiv 
\ell(\ell+1)C_\ell/2\pi$, the contribution to the $y$--DM covariance per logarithmic interval in $\ell$, which highlights the scales that dominate the signal.

The projected window functions $\tilde{y}_\ell(M,z)$ and $\widetilde{\DM}_\ell$ are the two-dimensional Fourier transforms of the halo profiles on the sky. For the tSZ:
\begin{equation}
  \tilde{y}_\ell =
  \frac{\sigma_T}{m_e c^2 D_A^2}
  \int_0^{r_\mathrm{max}} p_eP_\mathrm{th}(r|M,z)\,
  j_0\!\left(\ell r / D_A\right) 4\pi r^2\,dr,
  \label{eq:y_window}
\end{equation}
where $D_A$ is the angular diameter distance and $j_0(x) = \sin(x)/x$ is the zeroth-order spherical Bessel function. For the FRB DM:
\begin{equation}
  \widetilde{\DM}_\ell =
  \frac{1}{(1+z)\,D_A^2}
  \int_0^{r_\mathrm{max}} n_e(r|M,z)\,
  j_0\!\left(\ell r / D_A\right) 4\pi r^2\,dr,
  \label{eq:DM_window}
\end{equation}
where we integrate to $r_\mathrm{max} = 5 \times R_{200m}$, where both functions are converged. Both profiles are computed from the BP model for each halo mass $M$ and redshift $z$.

We calculate the angular cross power spectrum with a mass integral over $10^{13} - 10^{16} M_{\odot}$ and a redshift integral over $z \in [0.01, 2.0]$. We do not include an FRB redshift source kernel in our calculations, which effectively assumes all FRB sources are at $z=2$ (see Section~\ref{sec:frb_sys} for the impact of a realistic source kernel). The lower limit on the mass integral is set by the regime in which the BP model is validated \citep{Lau2025}.
 
The angular correlation function $w(\theta)$ is related to the power spectrum by
\begin{equation}
  w(\theta) = \sum_\ell \frac{2\ell+1}{4\pi}\,\Cl\,P_\ell(\cos\theta),
  \label{eq:w_theta}
\end{equation}
where $P_\ell$ are Legendre polynomials. We present results primarily in $D_\ell$ space, but compare to observational results in $w(\theta)$ space.

\subsection{Observational Comparison and Statistical Inference} \label{sec:mcmc_met}

First, to assess the information content of the tSZ auto-spectrum ($C_\ell^{yy}$), the cross-spectrum ($\yDM$), and their combination ($C_\ell^{yy}$ + $\yDM$), we perform Fisher forecasts around the fiducial BP model. Derivatives $\partial C_\ell/\partial\theta_i$ are computed by five-point finite differences in each of the seven parameters of the MCMC analysis, and the per-multipole Gaussian covariance includes the $C_\ell^{yy}$--$\yDM$ cross-covariance induced by the shared $y$ field. At small scales, the covariance is dominated by shot noise rather than non-Gaussianity. We consider two regimes: an idealized, noise-free case where we only consider sample variance of both fields (Section~\ref{sec:fisher}), and a realistic case using the component-separated $y$-map noise of forthcoming CMB experiments and the FRB dispersion-measure shot noise expected from DSA-2000 (Section~\ref{sec:future}). Both analyses consider scales of $\ell \in [30, 8000]$. 

As a measure of degeneracy, we quote the conditional correlation:
\begin{equation}
    r_{\rm cond} = -F_{ij}/\sqrt{F_{ii}F_{jj}}, 
\end{equation}
computed from the pre-prior data Fisher matrix with all other parameters held fixed, where a value closer to 1 indicates complete degeneracy and a value closer to 0 indicates independence. Unlike $r_{\rm marg}$, which describes the degeneracy direction given the remaining parameter uncertainties, $r_{\rm cond}$ is a property of the pre-prior data Fisher matrix: it is sensitive to noise but independent of priors, marginalization choices, $f_{\rm sky}$, and binning.

We summarize the joint constraint with the figure of merit:
\begin{equation}
    \mathrm{FoM} \equiv [\det 
    \mathbf{\Sigma}(\epsilon_f, A_\mathrm{nt})]^{-1/2},
\end{equation}
the inverse area of the two-dimensional error ellipse, computed with $\epsilon_f$ in units of $10^{-6}$, where $\Sigma$ is the $2 \times 2$ 
marginalized covariance. Larger FoM values correspond to tighter joint constraints.

Second, to assess the constraining power of the recent observations of \citet{Takahashi2025} and \citet{Sharma_2026}, we perform a Markov Chain Monte Carlo (MCMC) analysis implemented with \texttt{emcee} \citep{ForemanMackey2013}, running $94$ walkers via \texttt{MPIPool} for parallelization across nodes. 

\begin{figure*}
    \centering
    \includegraphics[width=\linewidth]{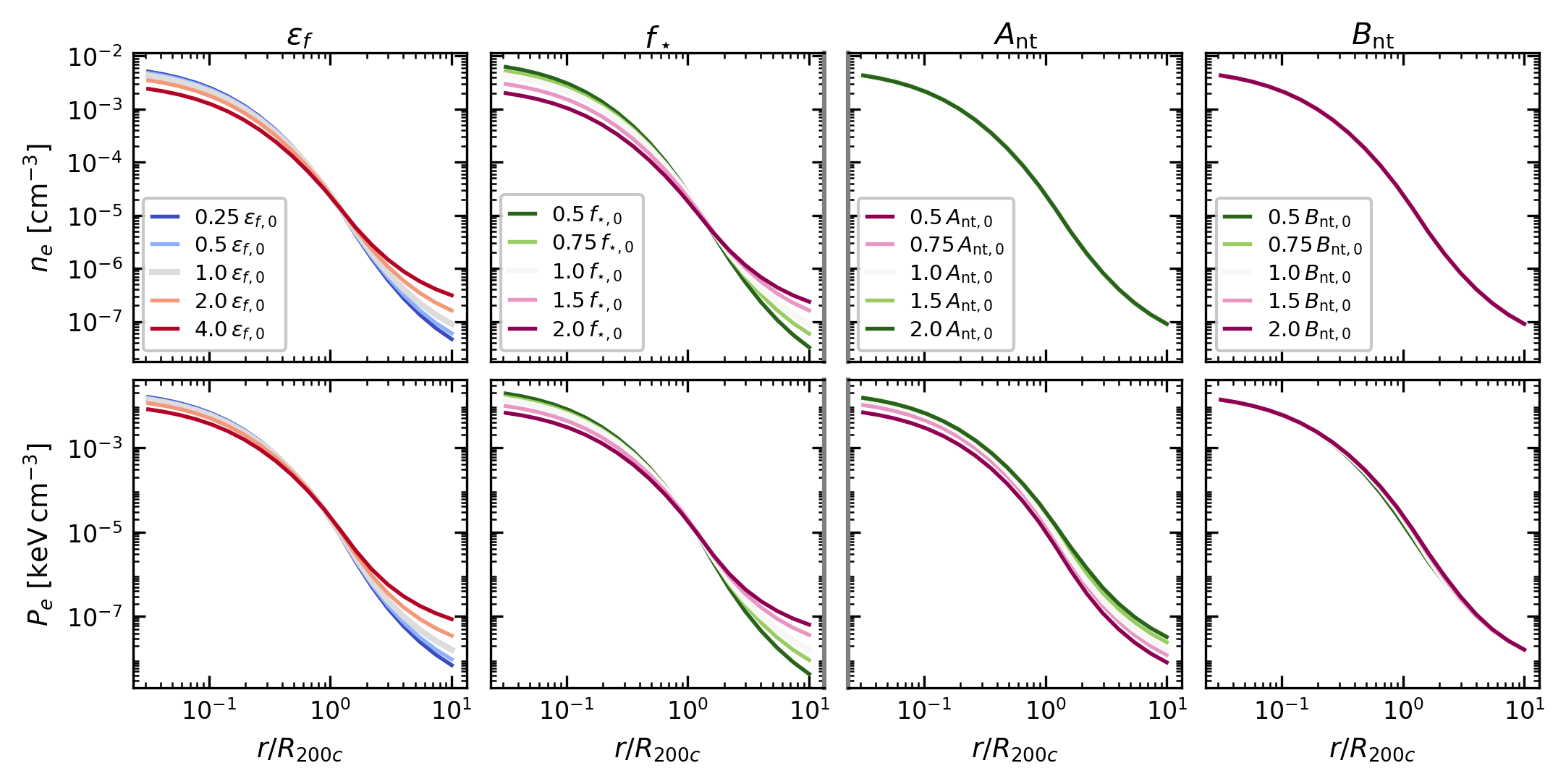}
    \caption{Sensitivity of the electron density (top row, $n_e$) and electron pressure (bottom row, $P_e$) halo radial profiles to four BP model parameters, shown as a function of radius $r/R_{200c}$ for a fixed halo of $M_h = 10^{14}\,M_\odot$ at $z = 0$. Each column varies one parameter: the feedback efficiency $\epsilon_f$, the stellar fraction $f_\star$, and the non-thermal pressure amplitude $A_{\rm nt}$ and transition scale $B_{\rm nt}$, by multiplicative factors of their fiducial values (see legends).}
    \label{fig:profiles_params}
\end{figure*}

The free parameters are five BP model parameters: the feedback efficiency $\epsilon_f$, the stellar fraction normalization and mass slope $(f_\star, S_\star)$, the non-thermal pressure amplitude $A_{\rm nt}$, and the pressure boundary radius $P_{\rm bound}$ (in units of the splashback radius). We also vary two cosmological parameters: $\Omega_m$ and $\sigma_8$. The $f_\star$, $S_\star$, and $P_{\rm bound}$ parameters carry Gaussian priors from \cite{Flender_2017} and \cite{Aung2021}, where $f_\star = 0.026 \pm 0.003$, $S_\star = 0.12 \pm 0.10$, and $P_{\rm bound} = 1.89 \pm 0.189$ (walled at [0.01, 2.5]). In addition, $\epsilon_f$ and $A_\mathrm{nt}$ carry flat priors, $\epsilon_f \in [0.001, 50]$ in units of $10^{-6}$ and $A_\mathrm{nt} \in [0.01, 0.99]$. The $\epsilon_f$ prior spans roughly five orders of magnitude and is deliberately uninformative; the one-sided limits reported in Section~\ref{sec:mcmc} should be read as bounded by this choice. We impose Gaussian priors $\Omega_m = 0.315 \pm 0.007$ and $\sigma_8 = 0.811 \pm 0.006$ motivated by CMB constraints.

In each likelihood evaluation, the theoretical cross-power spectrum $\yDM$ is computed on a sparse logarithmic $\ell$ grid, interpolated onto the full integer grid $\ell\in[2,10^4]$ (accurate to $<0.1\%$), and projected to $w(\theta)$ via Equation \ref{eq:w_theta}. The data span scales $\theta \sim 10'$--$200'$ for \citet{Sharma_2026} and $\theta \sim 1'$--$800'$ for \citet{Takahashi2025}, which correspond roughly to $\ell \sim 50 - 1100$ and $\ell \sim 10 - 11,000$, respectively. We exclude scales below $\theta_{min} = 10'$ for Planck observations and $\theta_{min} = 2'$ for ACT from the analysis, which correspond to the beam sizes of the detectors.

Convergence is assessed from the integrated autocorrelation time $\tau$, requiring $>100\,\tau$ steps and a stable $\tau$ estimate, and we discard $3\tau_{\max}$ as burn-in (mean acceptance $\simeq0.39$; effective sample sizes of order $10^5$ per parameter).

\section{Results}
\label{sec:results}

\subsection{Halo Profiles}

We begin by demonstrating the sensitivity of the electron density and electron pressure profiles to four of the key BP model parameters in Figure~\ref{fig:profiles_params} (for a halo with $M_h = 10^{14}\,M_\odot$ at $z = 0$). We vary the feedback strength ($\epsilon_f$), stellar fraction ($f_{\star}$), and non-thermal pressure ($A_{\rm nt}$ and $B_{\rm nt}$), by multiplying the fiducial value by various factors.

\begin{figure*}
    \centering
    \includegraphics[width=\linewidth]{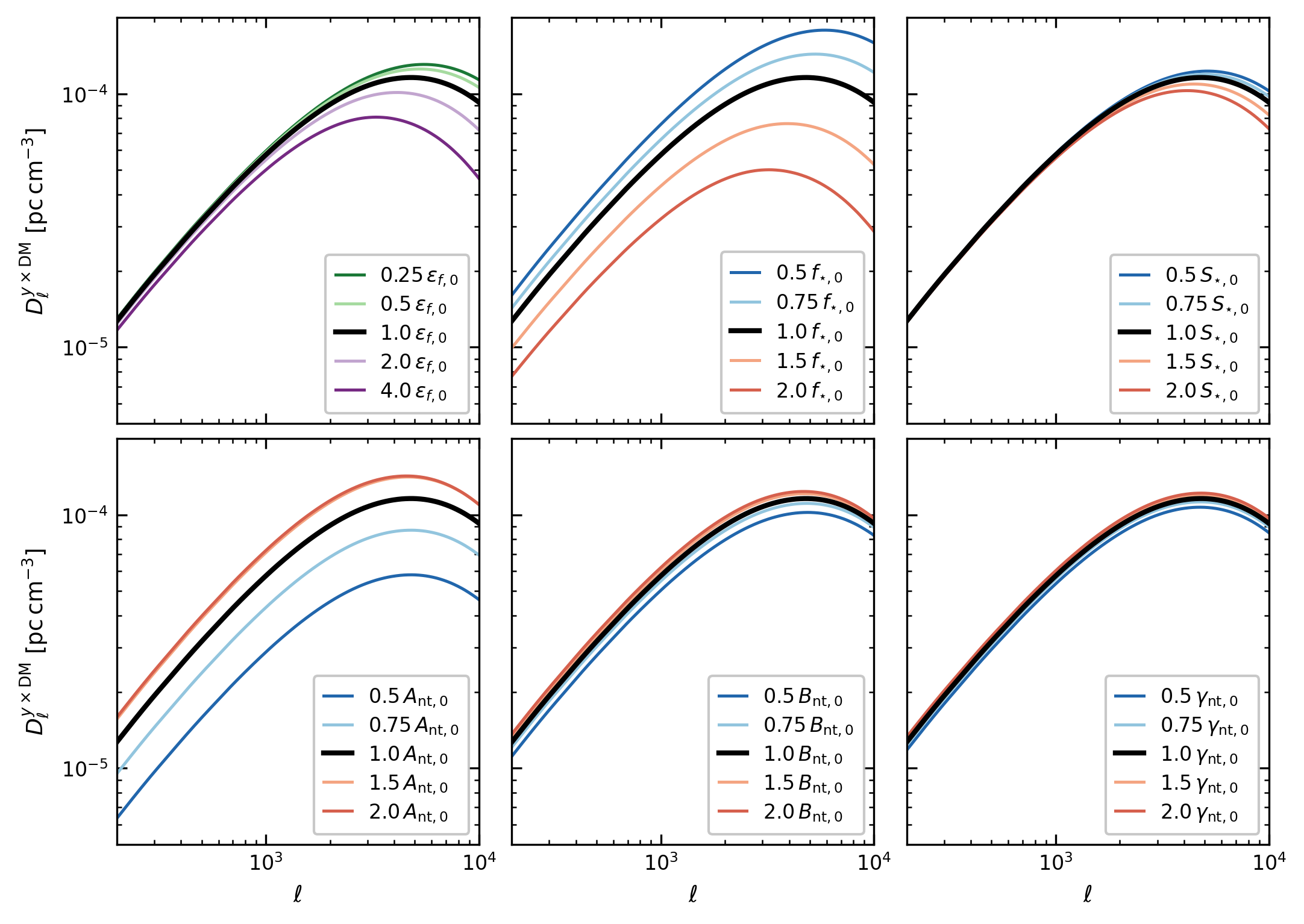}
    \caption{The tSZ$\,\times\,$FRB cross-power spectrum in the BP model as a function of multipole $\ell$, shown for variations in six key parameters. The top row varies the feedback- and stellar-related parameters (feedback efficiency $\epsilon_f$, stellar fraction $f_\star$, and its mass slope $S_\star$), and the bottom row varies the non-thermal pressure parameters (amplitude $A_{\rm nt}$, transition scale $B_{\rm nt}$, and slope $\gamma_{\rm nt}$); in each panel one parameter is scaled by multiplicative factors of its fiducial value (see legends), with the fiducial model shown as the solid black line.}
    \label{fig:spectra_varied}
\end{figure*}

As expected, the feedback and stellar parameters $\epsilon_f$ and $f_\star$ modulate both the electron density and pressure profiles, while $A_{\rm nt}$ redistributes the total pressure between thermal and non-thermal components without altering the overall gas distribution, leaving the density profiles unchanged. Varying $\epsilon_f$, in particular, produces a more distinctive effect, where stronger feedback evacuates the halo core and boosts the profiles at larger radii, as the gas is driven outward. This results in a redistribution of the radial profiles, not just their amplitude. The variation of $B_{\rm nt}$, which modulates the transition scale of the non-thermal pressure fraction, results in a very mild effect, only noticeable at scales near $R_{200c}$.

\subsection{The tSZ-FRB Cross-Power Spectrum} \label{sec:crosspower}

We present the angular cross-power spectra $D_\ell^{y \times \mathrm{DM}}$ as a function of six key BP model parameters in Figure~\ref{fig:spectra_varied}. The fiducial spectrum is shown in each panel for reference, with the black line. The two fields are positively correlated, with the signal peaking near $\ell \sim \mathrm{3000-5000}$ ($\theta \sim 2.2'-3.6'$), corresponding to the angular scales of the cluster interiors ($r < R_{\rm 500c}$). In the top row, we plot variations in parameters related to baryonic feedback and stellar properties: $\epsilon_f$, $f_{\star}$, and $S_{\star}$. In the bottom row, we plot variations in non-thermal pressure parameters: $A_{\rm nt}$, $B_{\rm nt}$, and $\gamma_{\rm nt}$.

First, we focus on the feedback and stellar parameters. Increasing $\epsilon_f$ suppresses the cross-power spectrum at all scales, as stronger feedback reduces both the electron density and pressure by evacuating gas from halos. In addition, it changes the peak power scale, with stronger feedback pushing it to larger scales (lower $\ell$ values). The amplitude of the stellar mass fraction, $f_{\star}$, modulates the amplitude of the cross power spectrum, notably at all scales of $\ell$ shown, not just the smaller scales. 

Higher $f_\star$ locks more baryons into stars, suppressing diffuse gas and reducing both signals. Within the \citet{Flender_2017} prior, the peak amplitude varies by only a few percent. $f_\star$ also shifts the peak scale, but less strongly than $\epsilon_f$. $S_{\star}$, the mass slope of the stellar mass fraction, modulates the amplitude of the smaller scales, with higher values suppressing power, similar to how $f_{\star}$ sets the fraction of baryons in stars versus gas, but to a less dramatic effect. $S_{\star}$ and $f_{\star}$ are degenerate given a measurement of the stellar-to-halo mass relation, as we see in Section~\ref{sec:fisher}.

Now, focusing on non-thermal pressure, we see that increasing $A_{\rm nt}$ enhances the spectrum, exclusively through the tSZ channel. Higher $A_{\rm nt}$ corresponds to a smaller fraction of non-thermal pressure, enhancing the thermal pressure at a fixed total pressure and thus boosting $y$, while leaving $\mathrm{DM}$ unchanged. The range of $A_{\rm nt}$ varied here corresponds to $f_{\rm nt} \in [0.096, 0.774]$ at large radii. The other two parameters, $B_{\rm nt}$ and $\gamma_{\rm nt}$, also slightly modulate the amplitude of the signal.

\begin{figure*}
    \centering
    \includegraphics[width=\linewidth]{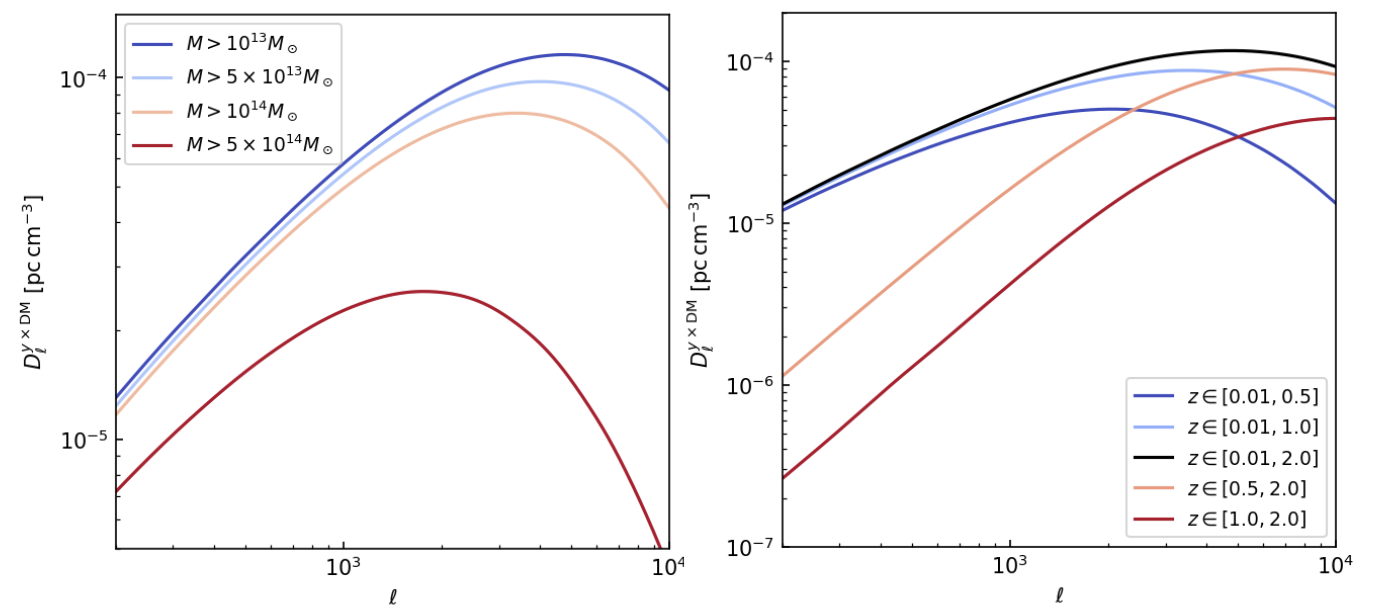}
    \caption{Sensitivity of the tSZ$\,\times\,$FRB cross-power spectrum $D_\ell^{y\times\mathrm{DM}}$ (in pc\,cm$^{-3}$) to the mass and redshift integration limits, as a function of multipole $\ell$. \textit{Left:} the lower limit of the mass integral is raised from $10^{13}$ to $5\times10^{14}\,M_\odot$ (see legend), with the upper limit fixed at $M = 10^{16}\,M_\odot$. \textit{Right:} the redshift integration window is varied (see legend); the near-overlap of $z\in[0.01,2.0]$ and $z\in[0.01,1.0]$ shows that most of the signal arises at $z < 1$.}
    \label{fig:sensitivity}
\end{figure*}

In Figure~\ref{fig:sensitivity} we examine which halo masses and redshifts dominate the tSZ$\,\times\,$FRB cross-power spectrum, by varying the limits of the mass and redshift integrals in Equations (\ref{eq:Cl_1h})--(\ref{eq:Cl_2h}). The left panel raises the lower limit of the mass integral while keeping the upper limit fixed at $M = 10^{16}\,M_\odot$. The curve for $M > 5\times10^{13}\,M_\odot$ show a 12.3\% drop in amplitude at $\ell = 3000$ compared to the $M > 10^{13}\,M_\odot$ curve, indicating that halos below $\sim 5\times10^{13}\,M_\odot$ do not contribute significantly at the scale of current observational precision. Raising the cut to $M > 5\times10^{14}\,M_\odot$ suppresses the signal substantially. The bulk of the cross-power therefore arises from group- to cluster-scale halos near $M \sim 10^{14}\,M_\odot$. The right panel varies the redshift integration window. The full $z \in [0.01, 2.0]$ result is closely tracked by $z \in [0.01, 1.0]$, indicating that the majority of the signal originates at $z < 1$, with only a modest contribution from $z > 1$. These sensitivities justify the mass and redshift integration ranges adopted in Section~\ref{sec:methods} and demonstrate that $\yDM$ probes feedback primarily in the group-to-cluster regime at low redshift, where the BP model is best validated against hydrodynamical simulations and X-ray observations.

\begin{table*}
    \centering
    \begin{tabular}{llccccc}
    \toprule
    Configuration & Observable & $\sigma(\epsilon_f)$ & $\sigma(A_{\rm nt})$ & $r_{\rm cond}$ & $r_{\rm marg}$ & Figure of Merit \\
    \midrule
    \multirow{3}{*}{CV-limited, Planck 2018 cosmology, no gas priors}
     & $C_\ell^{yy}$    & 31.6   & 0.465   & $+0.965$ & $+0.994$ & 0.63 \\
     & $\yDM$    & 16.2   & 0.560   & $+0.952$ & $+0.990$ & 0.79 \\
     & $C_\ell^{yy}$ + $\yDM$   & 0.197  & 0.0016  & $+0.081$ & $+0.741$ & $4.65\times10^{3}$ \\
    \midrule
    \multirow{3}{*}{CV-limited, Planck 2018 cosmology, gas priors}
     & $C_\ell^{yy}$    & 2.90   & 0.0143  & $+0.965$ & $+0.450$ & 27 \\
     & $\yDM$    & 1.98   & 0.0269  & $+0.952$ & $+0.257$ & 19 \\
     & $C_\ell^{yy}$ + $\yDM$   & 0.196  & 0.0016  & $+0.081$ & $+0.739$ & $4.69\times10^{3}$ \\
    \midrule
    \multirow{3}{*}{CV-limited, cosmology marginalized, gas priors}
     & $C_\ell^{yy}$    & 2.96   & 0.0200  & $+0.965$ & $+0.398$ & 18 \\
     & $\yDM$    & 2.02   & 0.0338  & $+0.952$ & $+0.166$ & 15 \\
     & $C_\ell^{yy}$ + $\yDM$   & 0.536  & 0.0054  & $+0.081$ & $-0.114$ & 346 \\
    \bottomrule
    \end{tabular}
    \caption{Cosmic-variance-limited (noise-free) Fisher forecast for the $\epsilon_f$--$A_\mathrm{nt}$ degeneracy from the tSZ auto-spectrum ($C_\ell^{yy}$), the cross-spectrum ($\yDM$), and their combination ($C_\ell^{yy}$+$\yDM$), for three prior and cosmology configurations. $\sigma(\epsilon_f)$ is in units of $10^{-6}$. The FoM is from the marginalized covariance and is computed with $\epsilon_f$ in units of $10^{-6}$.}
    \label{tab:fisher_ideal}
\end{table*}

\subsection{Breaking the Feedback and Non-Thermal Pressure Degeneracy} \label{sec:fisher}

\begin{figure*}
    \centering
    \includegraphics[width=\linewidth]{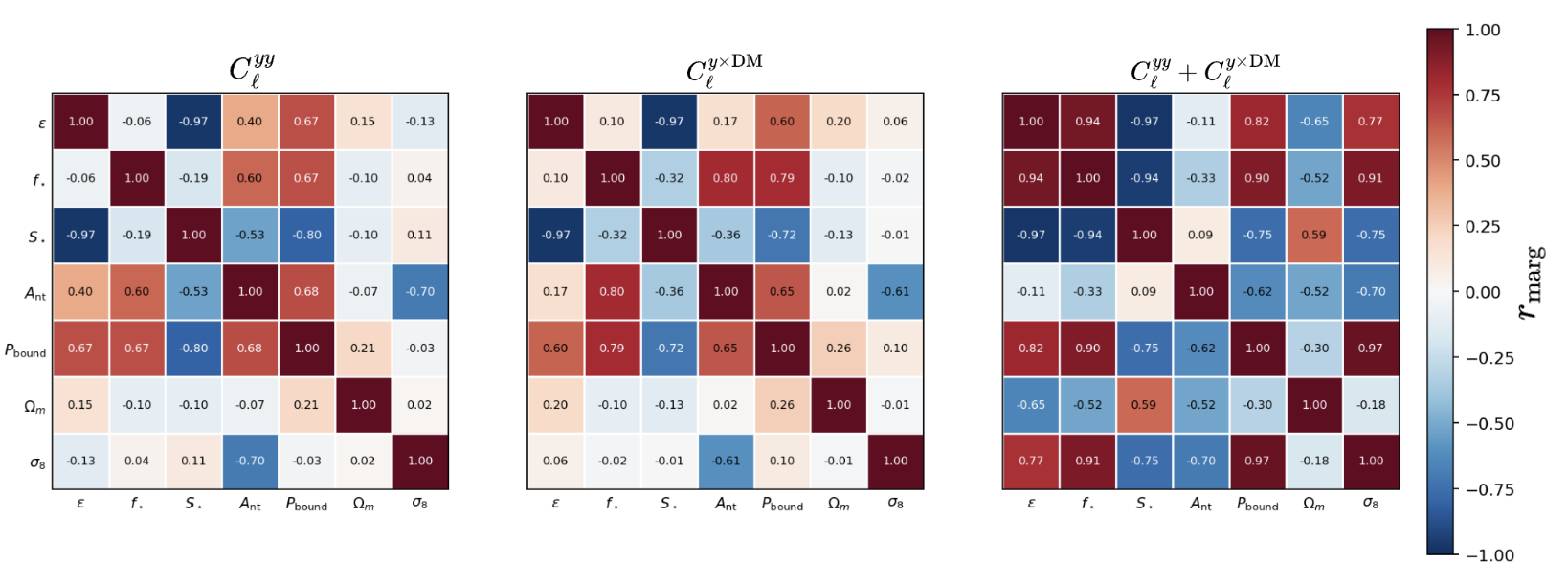}
    \caption{Marginalized correlation matrices of the BP model parameters ($\epsilon_f$, $f_\star$, $S_\star$, $A_\mathrm{nt}$, $P_\mathrm{bound}$) and the sampled cosmology ($\Omega_m$, $\sigma_8$), from the cosmic-variance-limited Fisher matrix with the priors of Section~\ref{sec:mcmc_met} applied, for the tSZ auto-spectrum ($C_\ell^{yy}$ ), the cross-spectrum ($\yDM$), and their combination. Red and blue denote positive and negative correlations; values near $\pm 1$ mark nearly degenerate pairs. Unlike $r_\mathrm{cond}$, these marginalized correlations depend on the prior and marginalization choices. In the joint constraint, as the $\epsilon_f$--$A_\mathrm{nt}$ degeneracy is reduced, the uncertainty shifts to the other correlations.}
    \label{fig:cor_mat}
\end{figure*}

The degeneracy-breaking power of joint analysis rests on the physical asymmetry between $\epsilon_f$, which modulates both the electron density and pressure profiles, and $A_{\rm nt}$, which rescales only the thermal pressure at fixed gas content. This causes the two response vectors to diverge between $C_\ell^{yy}$ and $\yDM$, as quantified by the Fisher analysis presented in Table~\ref{tab:fisher_ideal}. In every configuration $r_{\rm cond} \simeq 0.965$ for $C_\ell^{yy}$ and $\simeq 0.952$ for $\yDM$, indicating near-perfect alignment, and collapsing to $0.081$ in the combination. Because $r_{\rm cond}$ is a property of the pre-prior data Fisher matrix and is only sensitive to noise, this collapse is identical whether cosmology is fixed or marginalized and whether gas priors are applied. The marginalized uncertainties $\sigma(\epsilon_f)$ and $\sigma(A_{\rm nt})$ quantify the gain in constraining power. In the fully marginalized configuration, the joint $\epsilon_f$--$A_{\rm nt}$ figure of merit improves by a factor of $\sim$19 over the auto-spectrum alone, with $\sigma(\epsilon_f)[\times 10^{-6}]$ tightening from 2.96 to 0.54.

Figure~\ref{fig:cor_mat} presents the marginalized correlation matrices of the BP model and cosmological parameters for the tSZ auto-spectrum, the cross-spectrum, and their combination. Within either single spectrum, the strongest degeneracy is between $\epsilon_f$ and $S_\star$ ($r \simeq -0.97$), which trade off in setting the gas content of the group-scale halos that dominate the signal, while $\epsilon_f$--$A_\mathrm{nt}$ and $A_\mathrm{nt}$--$\sigma_8$ form the moderate amplitude-like degeneracies expected from the profile analysis above. $\epsilon_f$--$A_\mathrm{nt}$ highlights the degeneracy-breaking power directly, with $+0.40$ in $C_\ell^{yy}$, down to $+0.17$ in $\yDM$, and collapsing to $-0.11$ in the combination. 

We note that aside from $\epsilon_f$--$A_\mathrm{nt}$, the degeneracies in the joint panel appear to increase. With increased constraints on the $\epsilon_f$--$A_\mathrm{nt}$ direction, the uncertainty moves into combinations of the gas parameters and cosmology that also set the signal amplitude (e.g., $P_\mathrm{bound}$--$\sigma_8$ and $f_\star$--$\sigma_8$ correlations of $\simeq 0.9$), several of which strengthen relative to the single-probe panels. In the case of $\epsilon_f$--$S_\star$ in particular, this is the result of the data only constraining the total energy budget, not the energy efficiency itself. 

To identify which angular scales carry the degeneracy-breaking power, we repeat the joint Fisher analysis on cumulative multipole windows, retaining only $\ell\le\ell_{\rm cut}$ (growing the window from large scales) or only $\ell\ge\ell_{\rm cut}$ (growing from small scales), and recomputing $r_{\rm cond}$ and the figure of merit for each cut. We find that windows containing only small scales sit at the full-range floor for every cut (even $\ell\gtrsim5000$ alone yields $r_{\rm cond}\simeq0.08$ and retains a figure of merit within a factor of $\sim$4 of the full range), while windows restricted to large scales never approach it, retaining $r_{\rm cond}\gtrsim0.66$ for all $\ell_{\rm cut}\lesssim 1000$. The degeneracy-breaking power therefore lives in the one-halo regime of cluster interiors, $\ell\gtrsim3000$, where non-thermal pressure reshapes $\tilde y_\ell$ at fixed $\widetilde DM_\ell$ while feedback moves both window functions. This is consistent with the location of the signal peak (Section~\ref{sec:crosspower}) and has a direct consequence for observational instruments. To fully break the feedback and non-thermal pressure degeneracy, we need low $y$-map noise at $\ell\sim3000$--$8000$, not on large scales, which sets the ordering of the experiments considered in Section~\ref{sec:future}.

\subsection{Constraining the BP Model} \label{sec:mcmc}

\begin{table*}
\centering
\begin{tabular}{lcccc}
\toprule
 & Sharma+26 & \multicolumn{3}{c}{Takahashi+25} \\
 
\cmidrule(lr){2-2} \cmidrule(lr){3-5}
$y$-map & Planck NILC & ACT & Planck MILCA & Planck NILC \\

\midrule
\multicolumn{5}{c}{\textit{Goodness of fit}} \\
\hline
$N_\mathrm{data}$                & 15    & 11    & 8      & 8     \\
$\chi^2_\mathrm{H}$              & 4.1   & 4.8   & 18.7   & 6.7   \\
PTE                              & 0.997 & 0.94 & 0.02  & 0.57 \\

\addlinespace
\multicolumn{5}{c}{\textit{Feedback and non-thermal pressure}} \\
\hline

$\epsilon_f\,[10^{-6}]^{\,95\%}$   & $3.67$ & $2.55$ & $2.40$ & $2.59$ \\
$A_\mathrm{nt}$    & $0.67^{+0.22}_{-0.26}$ & $0.31^{+0.34}_{-0.21}$ & $0.45^{+0.36}_{-0.31}$ & $0.49^{+0.33}_{-0.32}$ \\
$A_\mathrm{nt}$  $\sigma_\mathrm{post}/\sigma_\mathrm{prior}$ & 0.72 & 0.76 & 0.99 & 0.97 \\
$r_\mathrm{P}$    & $-0.19$ & $+0.03$ & $-0.03$ & $-0.04$ \\
$r_\mathrm{S}$    & $-0.18$ & $+0.03$ & $-0.03$ & $-0.04$ \\

\addlinespace

\multicolumn{5}{c}{\textit{Prior-recovered parameters}} \\
\hline
$f_\star$     & $0.026\pm0.003$ & $0.026\pm0.003$ & $0.026\pm0.003$ & $0.026\pm0.003$ \\

$S_\star$     & $0.14\pm0.1$    & $0.11\pm0.1$   & $0.11\pm0.1$   & $0.12\pm0.1$ \\

$P_\mathrm{bound}$ & $1.8\pm0.2$  & $1.9\pm0.19$  & $1.9\pm0.18$ & $1.9\pm0.19$ \\

$\Omega_m$    & $0.32\pm0.007$  &  $0.31\pm0.007$ & $0.31\pm0.007$  & $0.32\pm0.007$ \\

$\sigma_8$    & $0.81\pm0.006$  & $0.81\pm0.006$ & $0.81\pm0.006$ & $0.81\pm0.006$ \\

\bottomrule
\end{tabular}
\caption{Marginalized parameter constraints and fit quality for the four measurements from \citet{Sharma_2026} and \citet{Takahashi2025}, each fit independently. Parameter values are marginalized medians with 68\% credible intervals, except for $\epsilon_f$ where we report the one-sided $95\%$ lower limit, as it is prior-dominated, with the posterior rising gently towards the upper boundary of the 50$\times 10^{-6}$ prior. $\sigma_\mathrm{post}/\sigma_\mathrm{prior}$ is the $A_\mathrm{nt}$ posterior-to-prior width ratio (where $\sigma_\mathrm{post}$ is the chain standard deviation), and $r_\mathrm{P}$ and $r_\mathrm{S}$ are the Pearson and Spearman correlations of $(\epsilon_f, A_\mathrm{nt})$ from the chain. $\chi^2_\mathrm{H}$ is the Hartlap-debiased \citep{Hartlap_2007} $\chi^2$, with the probability to exceed (PTE) computed for $N_\mathrm{data}$ degrees of freedom. }

\label{tab:mcmc}
\end{table*}

We compare the BP model to the two existing detections: \citet{Takahashi2025}, who measured the cross-correlation of the DMs of 133 localized FRBs and Planck NILC (Needlet Internal Linear Combination) and MILCA (Modified Internal Linear Combination Algorithm) $y$-maps ($4\sigma$) and ACT ($1.5\sigma$), and \citet{Sharma_2026}, who achieved a $3.8\sigma$ detection with 3455 unlocalized CHIME FRBs and the Planck NILC $y$-map. The broader angular coverage of the CHIME sample provides additional leverage on the two-halo term, which is more sensitive to the large-scale gas distribution.  Within the \cite{Takahashi2025} measurement, the two Planck reconstructions share the same 71 FRBs and the same Planck data, while the ACT measurement uses an independent footprint with 31 FRBs and probes smaller scales (leading to the lower-significance measurement). We therefore fit four data vectors independently.

\begin{figure*}
    \centering
    \includegraphics[width=\linewidth]{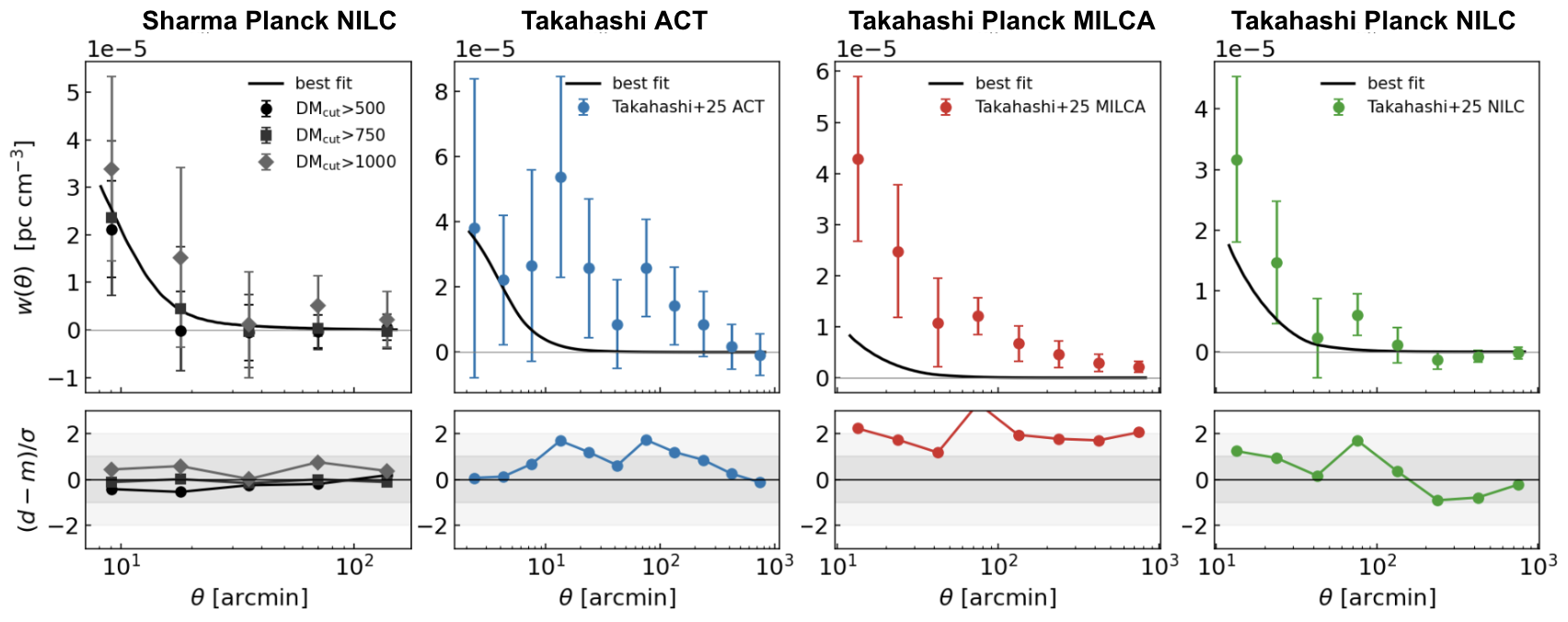}
    \caption{Measured tSZ$\times$FRB angular correlation functions with the best-fit BP model for each of the four independent fits. \emph{Top panels:} $w(\theta)$ for \cite{Sharma_2026} (three DM cuts, ${\rm DM}_{\rm cut}=500$, 750, $1000\,{\rm pc\,cm^{-3}}$) and the three \cite{Takahashi2025} $y$-map measurements (ACT, Planck MILCA, Planck NILC), each with the maximum-likelihood model from its own chain (black curve). Error bars are the square roots of the diagonal of each survey's jackknife covariance. The fits use the full covariance. The Planck-based panels begin at $\theta_{\rm min}=10'$ ($2'$ for ACT), below which the $y$-map beam suppresses the data relative to the un-beamed model. \emph{Bottom panels:} standardized residuals $(d-m)/\sigma$. Shaded bands mark $\pm1\sigma$ and $\pm2\sigma$.}
    \label{fig:BP_fit}
\end{figure*}

For each survey, we perform an MCMC analysis (as described in Section~\ref{sec:mcmc_met}). In Table~\ref{tab:mcmc} we present the marginalized constraints for all four fits, as well as the goodness of fit metrics. Figure~\ref{fig:BP_fit} shows each measurement with its best-fit model (the maximum-likelihood sample of each chain) and standardized residuals. Three of the four fits are good (PTE $=0.57$--$0.997$).

The MILCA fit is noticeably poor ($\chi^2_{\rm H}=18.7/8$, PTE $=0.02$) with a coherently positive residual, while NILC (built from the same Planck data and the same 71 FRBs) fits well comparatively ($\chi^2_{\rm H}=6.7/8$). A direct comparison shows that the two data vectors agree within their errors bin by bin, but MILCA lies above NILC in every bin, with a coherent additive offset of up to $\sim2\sigma$ per bin, producing a systematic rather than statistical excess. Because the only difference between the two measurements is the component-separation weighting, this offset is a foreground residual, consistent with MILCA's known greater large-scale contamination \citep{MILCA_2016}. A decaying halo-model template cannot absorb an additive pedestal, and MILCA's stronger bin-to-bin correlations (median off-diagonal $+0.39$ versus $+0.12$ for NILC) concentrate the resulting $\chi^2$ penalty. A leave-out test confirms the contamination is broadband rather than confined to large scales. Excluding the largest-angle bins never restores an acceptable fit (reduced $\chi^2\geq2.4$ for all $\theta_{\max}$ cuts, versus $0.9$--$1.1$ for NILC under the same cuts). We therefore treat NILC as the primary Planck measurement and MILCA as a systematic cross-check, illustrating that component-separation residuals matter at the $\sim2\sigma$ level for this signal.

\begin{figure*}
    \centering
    \includegraphics[width=\linewidth]{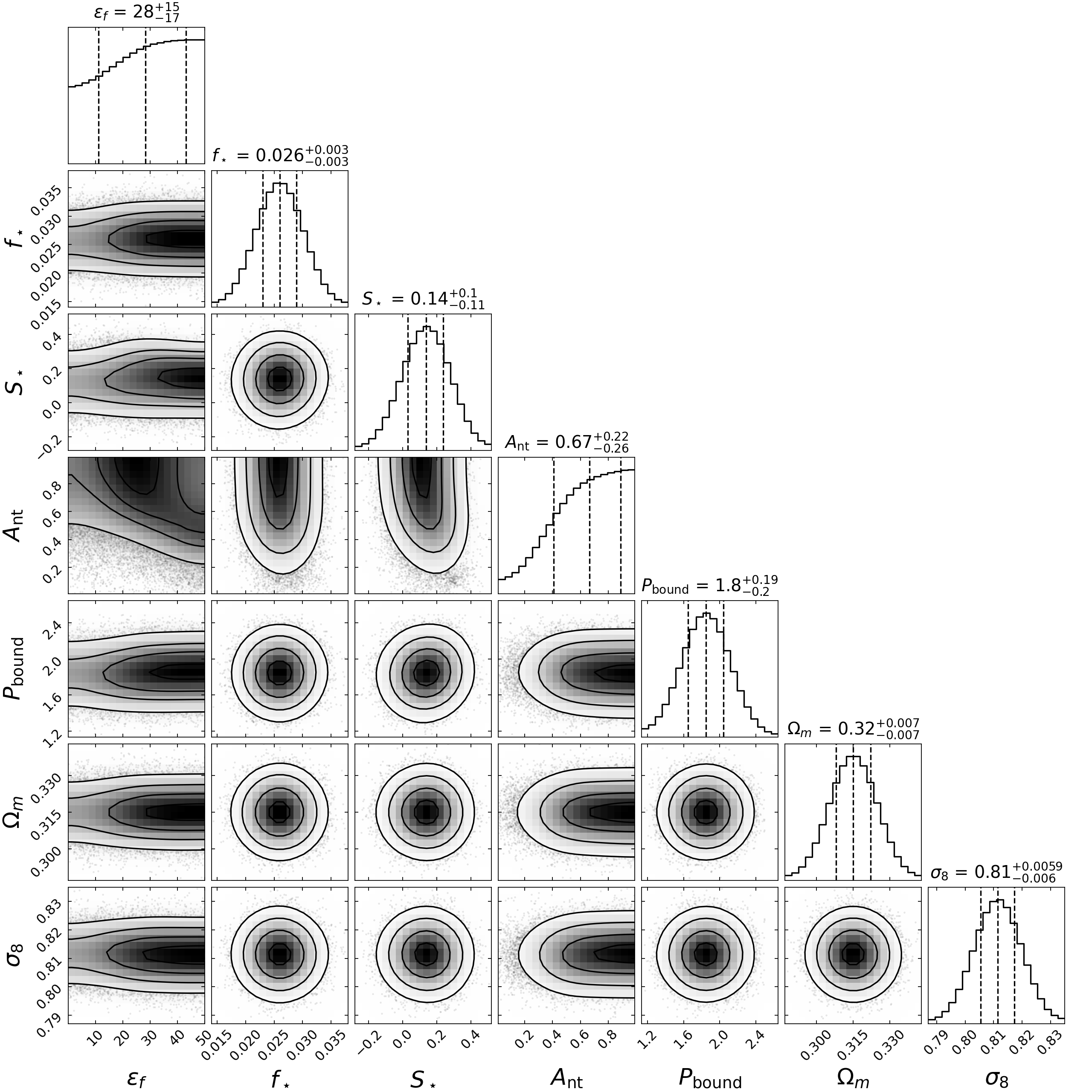}
    \caption{The full posterior distributions for five BP model parameters ($\epsilon_f$, $f_\star$, $S_\star$, $A_{\rm nt}$, and $P_{\rm bound}$) and two cosmological parameters ($\Omega_m$ and $\sigma_8$) from the MCMC fit to the tSZ$\,\times\,$DM measurements of \citet{Sharma_2026}. Contours mark the 68\%, 95\%, and 99\% credible regions, and dashed lines in the marginalized 1D distributions indicate the median and 68\% credible interval, quoted above each column. $\epsilon_f$ is rail-limited so we report the lower limit at 95\% confidence in Table~\ref{tab:mcmc}.}
    \label{fig:mcmc_sharma}
\end{figure*}

Figure~\ref{fig:mcmc_sharma} shows the full posteriors for the \citet{Sharma_2026} fit. We find that in every fit the posterior of $\epsilon_f$ rises gently toward the upper prior boundary of $50 \times 10^{-6}$, set by the limits of the BP model. We therefore report a one-sided lower limit at 95\% confidence, $\epsilon_f^{\,95\%}$, rather than the median value. We measure $\epsilon_f^{\,95\%} > 3.67 \times 10^{-6}$ for the \citet{Sharma_2026} measurement, and $\epsilon_f^{\,95\%} > 2.55 \times 10^{-6}$, $2.40 \times 10^{-6}$, and $2.59 \times 10^{-6}$ for the \citet{Takahashi2025} ACT, Planck MILCA, and Planck NILC measurements, respectively. The fiducial $\epsilon_f = 3.97 \times 10^{-6}$ of the BP model, calibrated to X-ray data, lies above these lower limits, and all four fits disfavor a weaker feedback scenario. We caution, however, that because the posterior rails against the boundary, these limits are quoted within, and depend on, the adopted prior range. 

$A_\mathrm{nt}$ also exhibits a mildly rising posterior and approximately recovers its prior ($\sigma_{\rm post}/\sigma_{\rm prior} = 0.72$--$0.99$ against the flat $[0.01, 0.99]$ prior). $A_{\rm nt}$ is therefore unconstrained by current data. 

The $\epsilon_f$--$A_\mathrm{nt}$ plane quantifies the current-data degeneracy directly from the chains. The highest-S/N measurement \citep{Sharma_2026} shows a clear anti-correlation ridge ($r = -0.19$). The orientation is initially surprising, since increasing $\epsilon_f$ suppresses the cross-spectrum near its peak while increasing $A_\mathrm{nt}$ raises it, so amplitude compensation alone would give a positively sloped ridge, as in the cosmic-variance-limited forecast of Section~\ref{sec:fisher}. We discuss this in Section~\ref{sec:disc_int}. The \citet{Takahashi2025} maps are consistent with, but do not independently detect, this ridge ($r = +0.03$, $-0.03$, $-0.04$ for ACT, MILCA, NILC).

The stellar parameters $(f_\star, S_\star)$ remain prior-dominated, since the cross-power spectrum has limited sensitivity to the stellar mass fraction at the halo masses and angular scales probed. The gas boundary also recovers the prior $P_{\rm bound}=1.8\pm0.2$.

\subsection{Constraints with Future Surveys} \label{sec:future}

We now quantify the constraining power of realistic surveys expected within the next few years. For the $y$-map we adopt the Compton-$y$ noise $N_\ell^{yy}$ of \cite{Raghunathan_2022} (harmonic-space ILC residual). For the FRB field, we use the white shot-noise model of \citet{Sharma_2026c},
\begin{equation}
\begin{aligned}
N^{\rm DM\,DM} &= \frac{\sigma_{\rm field}^{2}}{\bar n}
   + \frac{\sigma_{\rm host}^{2}}{(1+z)^{2}\,\bar n} \\
&= \frac{1}{2\pi\bar n}\int \ell\, C_\ell^{\rm DM\,DM}\, d\ell
   + \frac{\sigma_{\rm host}^{2}}{(1+z)^{2}\,\bar n}
\end{aligned}
\end{equation}
with $\sigma_{\rm host} = 150\,{\rm pc\,cm^{-3}}$. The DSA-2000 five-year yield ($5\times10^{4}$ FRBs, $33{,}000\,{\rm deg^{2}}$, ${\rm dec} > -30^{\circ}$) gives $N^{\rm DM\,DM} = 4.5\,({\rm pc\,cm^{-3}})^{2}\,{\rm sr}$. SO and CMB-HD overlap this footprint at $f_{\rm sky} \simeq 0.3$; SPT does not. We forecast a sensitivity ladder at fixed $f_{\rm sky}$ varying only $N_\ell^{yy}$.

\begin{table}
\centering
\begin{tabular}{lccccc}
\toprule
Experiment & $\sigma(\epsilon_f)$ & $\sigma(A_{\rm nt})$ & $r_{\rm cond}$ & $r_{\rm marg}$ & FoM \\
\midrule
SO-Baseline & 4.27 & 0.0172 & $+0.973$ & $+0.282$ & 14.2 \\
SO-Goal     & 4.17 & 0.0167 & $+0.974$ & $+0.296$ & 15.1 \\
CMB-HD      & 2.99 & 0.0105 & $+0.974$ & $+0.376$ & 34.2 \\
\bottomrule
\end{tabular}
\caption{Fisher forecast at realistic survey noise for the ladder of CMB experiments whose footprints overlap DSA-2000 ($f_{\rm sky}\simeq0.3$ with nominal $5\times10^{4}$ localized FRBs, $\sigma_{\rm host}=150\,{\rm pc\,cm^{-3}}$), with CMB priors and cosmology marginalized and joint ($C_\ell^{yy}$ + $\yDM$) constraints. $\sigma(\epsilon_f)$ is in units of $10^{-6}$. The FoM is from the marginalized covariance and is computed with $\epsilon_f$ in units of $10^{-6}$.}
\label{tab:ladder}
\end{table}
 
Table~\ref{tab:ladder} summarizes the sensitivity ladder: the $\epsilon_f$--$A_{\rm nt}$ figure of merit increases from 14.2 (SO-Baseline) to 34.2 (CMB-HD), with $\sigma(A_{\rm nt}) = 0.0105$ and $\sigma(\epsilon_f) = 2.99\times10^{-6}$ at CMB-HD --- a factor of $\sim$27 improvement over the uninformative prior ($\sigma_{\rm prior} = 0.283$). Relative to the fiducial $A_{\rm nt} = 0.452$, this corresponds to fractional uncertainties of 3.8\% and 2.3\% for SO-Baseline and CMB-HD, respectively, though these do not account for the additional scatter sources discussed in Section~\ref{sec:disc_lim}.

The CMB-HD $y$-map is signal-dominated ($N_\ell^{yy}/C_\ell^{yy} < 1$) for $\ell \gtrsim 750$, reaching $N/C \simeq 0.1$ at $\ell \sim 7000$, so the one-halo window that carries the degeneracy-breaking power (Section~\ref{sec:fisher}) is essentially noise-free $y$. However, $r_{\rm cond} \simeq 0.97$ throughout, indicating near-perfect alignment of the $\epsilon_f$ and $A_{\rm nt}$ responses, confirming that CMB sensitivity alone cannot break the degeneracy.

The DM field is shot-noise-dominated at every multipole: $N^{\rm DM\,DM}/C_\ell^{\rm DM\,DM}\simeq35$ at $\ell=30$, rising to $665$--$3370$ at $\ell\gtrsim3000$. This means that the cross-spectrum is DM-noise-limited in exactly the window where the $\epsilon_f$ and $A_{\rm nt}$ responses decouple. The joint constraint is consequently dominated by the auto-spectrum, where $r_{\rm cond}\simeq0.97$. The limiting factor is therefore not the CMB sensitivity but the FRB survey constraints.

To reach a signal-dominated DM field on one-halo scales requires $N_{\rm FRB} \gtrsim (3.3$--$8.2)\times10^{7}$, three orders of magnitude above the nominal DSA-2000 yield. Even $10^{7}$ localized FRBs leave $r_{\rm cond} = 0.82$ in CMB-HD noise. Fully breaking this degeneracy therefore lies beyond foreseeable surveys unless the effective per-burst scatter is substantially reduced and the field variance is modeled accurately. 

Nevertheless, eliminating the correlation is not a prerequisite for constraining both parameters. $r_{\rm cond}$ measures the survival of a correlated direction, not the loss of constraining power. In the joint analysis, both parameters are individually well measured, and the joint constraint on $\epsilon_f$ and $A_{\rm nt}$ is expected to tighten by a factor of $\sim 1.6$ over the coming decade, a meaningful step toward quantifying the total non-thermal pressure budget in groups and clusters.

\section{Discussion}
\label{sec:disc}

The results of Section~\ref{sec:results} establish the degeneracy-breaking potential of the tSZ$\times$FRB cross-correlation. Here we interpret the physical origin of this signal and assess its robustness.

\subsection{Physical Interpretation}
\label{sec:disc_int}

Although the BP model parameterizes $\epsilon_f$ and $A_{\rm nt}$ independently, the thermal pressure profile depends on the feedback energy budget, so any physical correlation between the two will still emerge in measurements. The asymmetric response of $C_\ell^{y{\rm DM}}$ to $\epsilon_f$ and $A_{\rm nt}$ across angular scales is what allows the two processes to be disentangled. The concentration of degeneracy-breaking power in the one-halo regime reflects that the pressure-to-density ratio is most sensitive to feedback-driven thermodynamic changes at the scales where both signals peak. This is distinct from the well-known dominance of non-thermal pressure in cluster outskirts and the large-scale sensitivity of the tSZ$\times$FRB signal. XRISM currently probes gas motions directly within and around the cluster core ($r \lesssim 0.4\,R_{500c}$) \citep{xrism_comparison}, and tSZ fluctuation analyses provide indirect constraints at larger radii but are limited to nearby systems such as the Coma cluster \citep[e.g.][]{Khatri_2016}. The tSZ$\times$FRB cross-correlation offers complementary constraints on non-thermal pressure across a broader range of cluster regions, with the statistical reach of large FRB samples.

This scale dependence explains the orientation of the degeneracy ridge in the current MCMC fits (Section~\ref{sec:mcmc}). On the large angular scales probed by the present measurements, stronger feedback moves gas to large radii, increasing $w(\theta)$ in the same direction as increased thermal support. The $r = +0.03$ value for ACT reflects its sensitivity to smaller scales than Planck, approaching the regime where the two responses begin to de-align. Current data therefore constrain a scale-limited combination of feedback and non-thermal pressure, not the small-scale signal that carries the degeneracy-breaking power.

With these points and the caveats of Section~\ref{sec:disc_lim} in mind, the fits to current observations consistently disfavor weaker-than-fiducial feedback, consistent with the \citet{Sharma_2026} fits using the BCEmu baryon model, which favor a moderate feedback scenario with the same tSZ$\times$FRB signal and disfavor weak feedback with the soft X-ray background $\times$ FRB cross-correlation. While several recent observations favor stronger-than-simulated feedback \citep[e.g.][]{Bigwood_2024, Siegal_Feedback_2026}, others support a more moderate scenario \citep[e.g.][]{Sharma_2026a}. Current tSZ$\times$FRB measurements lack the precision to resolve this tension but will provide important context as sensitivity improves.

\subsection{Caveats and Limitations}
\label{sec:disc_lim}

\subsubsection{Limitations of the Baryon Pasting Model}

The BP model captures the essential physics of gas thermodynamics and feedback through physically motivated profiles, but necessarily makes several simplifying assumptions.

The smooth, spherically symmetric profiles assigned by the BP model break down during active mergers and in the presence of strong substructure. However, as shown in Section~\ref{sec:fisher} and Figure~\ref{fig:sensitivity}, the cross-power spectrum is dominated by the inner regions of massive halos at $z < 1$, where the assumptions of the smooth profile are best justified, and the BP model has been validated against hydrodynamical simulations and X-ray observations \citep{Lau2025}. An aspherical gas model is implemented in \citet{Lau2025} and modifies the selection function of individual halos by making elongated halos appear brighter or fainter depending on their orientation. However, at the map level the random orientations of halos average out in the ensemble, so asphericity adds scatter around the mean profile without biasing the mean cross-power; we therefore neglect it here.

Our halo model assigns a single deterministic gas profile to each halo at fixed mass and redshift, neglecting halo-to-halo scatter from differences in formation history, dynamical state, and feedback. \citet{Lau2025} showed that this halo-to-halo variance has non-negligible effects on the auto- and cross-power spectra of multiple observables, introducing bias in both cosmological and astrophysical parameter 
inference. Gas-rich halos are simultaneously denser and more pressurized, so this scatter introduces a positive density--pressure covariance in the one-halo term that our mean-profile calculation omits, biasing the predicted amplitude slightly low at small scales. Quantifying this effect for the tSZ$\times$FRB cross-correlation is left to future work.

Finally, in the outer regions of massive halos, the electron--ion equilibration time scale can exceed the dynamical time, leading to $T_e < T_i$ \citep{Rudd2009, Avestruz2015}. The single-fluid approximation $T_e = T_i$ adopted here may slightly overestimate $y$ in the halo outskirts, at the $\sim$10\% level for the mass and redshift range that dominates the signal.

\subsubsection{FRB Systematics}
\label{sec:frb_sys}

Throughout this work, we place all FRBs at $z = 2$, yielding the maximum-amplitude case. In practice, the \citet{Sharma_2026} DM-threshold samples ($\mathrm{DM}_{\rm cut} = 500$, 750, and $1000\,\mathrm{pc\,cm^{-3}}$) and the localized \citet{Takahashi2025} sample (mean $z = 0.3$) each carry their own redshift distributions. This does not affect conclusions drawn from differential responses of the cross-power spectrum, but does shift the MCMC results. A realistic source kernel suppresses $w(\theta)$ by 16--45\% and shifts the lower limit of $\epsilon_f$ by $\sim$5--22\%, well within the spread across the four datasets.

The FRB DM receives contributions from the Milky Way, the host galaxy, the IGM, and hot halo gas of intervening halos (Equation~\ref{eq:dmbreakdown}). The MW term is uncorrelated with the $y$-field and adds noise but not signal. The IGM term adds a correlated component sensitive to the large-scale baryon distribution rather than the gas physics of individual halos. Our halo-model approach models only the halo contribution explicitly. In our realistic forecasts we include a host variance $\sigma_{\rm DM,host} = 150\,\mathrm{pc\,cm^{-3}}$ and a correlated IGM foreground, which together set a floor on the FRB constraining power.

The host contribution $\mathrm{DM_{host}}/(1+z)$ is uncorrelated with the foreground $y$-field and adds variance rather than bias \citep[backlight argument of][]{Sharma_2026}. The exception is when the FRB host is itself a massive halo, whose gas contributes to both $\mathrm{DM_{host}}$ and the host's Compton-$y$ signal, producing a correlated term that can dominate at $\theta \lesssim 10'$ \citep{Takahashi2025}. This self-contribution must be removed before comparison with our model, as in the cluster-FRB exclusion of \citet{Takahashi2025}, but is negligible on the scales that dominate our signal.

\subsubsection{tSZ Systematics}

The tSZ signal is susceptible to contamination from Galactic thermal dust and the CIB \citep{Mroczkowski_2019, Chiang_2020}. Notably, \citet{Sharma_2026} detects an FRB $\times$ CIB correlation at significance comparable to the tSZ$\times$FRB signal. We refer the reader to \citet{Takahashi2025} and \citet{Sharma_2026} for details on how these were treated in the observations. Figure~\ref{fig:sensitivity} shows that most of our signal originates at $z < 1$, and our $z = 2$ source kernel further underpredicts this lower-redshift contribution. This is below the $z=1-3$ range where the CIB peaks \citep{Mroczkowski_2019}. Taken together with the fact that this CIB contamination is not squared, we expect tSZ$\times$FRB to be less sensitive to the CIB systematics than the tSZ autocorrelation. However, CIB contamination in the \textit{Planck} NILC (Needlet Internal Linear Combination) $y$-map is concentrated at small scales, precisely where the constraining power resides. Precision constraints will therefore require careful CIB mitigation, either via multi-frequency ILC with CIB deprojection or instruments with broader frequency coverage \citep[e.g.][]{McCarthy_2024, SO_2019}.

\subsubsection{Total Effect}

None of these systematics is expected to alter our conclusions at current sensitivity. Each affects the predicted amplitude at the tens-of-percent level or below, well within the current measurement uncertainties, and our central result rests on the differential response of $\yDM$ to $\epsilon_f$ and $A_{\rm nt}$ rather than its absolute normalization. Of these, halo-to-halo scatter, the FRB source kernel, and CIB contamination are most likely to require explicit treatment before precision claims at the $\sim$2--4\% level on $A_{\rm nt}$ can be made. Quantitative assessment of all systematics is deferred to future work.

\subsection{Outlook}
\label{sec:disc_fut}

Section~\ref{sec:future} quantifies the gain in joint $\epsilon_f$--$A_{\rm nt}$ constraining power from the growth in localized FRB samples and $y$-map sensitivity expected over the coming decade. Tomographic measurements, binning FRBs in redshift or using DM as a redshift proxy, could further track the thermodynamic evolution of halo gas over cosmic time. With redshift-dependent feedback, this offers a route to constraining the history of AGN feedback, complementing individual FRB--CGM measurements \citep{Connor2022, Wu2023, Baptista2024, Medlock_2024}, FRB constraints on the matter power spectrum \citep{Medlock_2025b}, and cross-correlations with galaxy catalogs to decompose the signal by halo mass and environment \citep{Medlock_2025a}.

The Baryon Pasting framework predicts a wide range of observables from a well-defined set of physically motivated parameters. The differentiability of \texttt{diffgas} and the BP framework's existing applications to X-ray emission \citep{Lau2025} and the kSZ effect make a self-consistent multi-wavelength analysis that combines $\yDM$ with the DM auto-spectrum, as well as X-ray, kSZ, and weak lensing cross-correlations, a natural next step toward breaking the residual $\epsilon_f$--$S_\star$ and amplitude--cosmology degeneracies identified in Section~\ref{sec:fisher}.

\section{Conclusions}
\label{sec:conc}

We have shown that the tSZ$\times$FRB cross-correlation can disentangle feedback-driven gas ejection and non-thermal pressure support, two processes that are degenerate in the tSZ auto-power spectrum alone. Feedback efficiency and non-thermal pressure amplitude imprint distinct scale-dependent signatures across $C_\ell^{yy}$ and $\yDM$ that are nearly perfectly degenerate within any single probe ($r_{\rm cond} \simeq 0.96$) but separable in their combination. Our main findings are as follows.

\begin{enumerate}

\item In the noise-free limit, combining $C_\ell^{yy}$ and $\yDM$ collapses $r_{\rm cond}$ from $\simeq 0.96$ to $0.08$ (a 12-fold reduction), improving the joint figure of merit by a factor of $\sim$19. The degeneracy-breaking power is concentrated in the one-halo regime at $\ell \gtrsim 3000$.

\item The fiducial BP model is consistent with current detections \citep{Takahashi2025, Sharma_2026}, with the signal dominated by halos near $M \sim 10^{14}\,M_\odot$ at $z < 1$, in the group-to-cluster regime.

\item Current data trace the $\epsilon_f$--$A_{\rm nt}$ degeneracy but cannot break it. Weak feedback is disfavored ($\epsilon_f^{95\%} \gtrsim 2.4$--$3.7 \times 10^{-6}$), while $A_{\rm nt}$ recovers its prior across all four datasets, confirming that non-thermal pressure support remains unconstrained at current sensitivity.

\item DSA-2000 combined with CMB-HD will constrain $A_{\rm nt}$ to $2.3\%$ fractional precision and improve the joint figure of merit by a factor of $\sim$27 over the uninformative prior. Fully breaking the degeneracy requires $N_{\rm FRB} \gtrsim 10^7$ localized bursts, a clear target for the next generation of FRB facilities.

\end{enumerate}

As FRB censuses grow and CMB experiments push toward higher sensitivity, the tSZ$\times$FRB cross-correlation will evolve from a detection-level measurement into a precision constraint on the thermodynamic state of group and cluster gas, with direct implications for hydrostatic mass bias corrections in eROSITA, SO, and CMB-HD cluster samples. Looking further ahead, combining $\yDM$ with X-ray, kSZ, and weak lensing cross-correlations within the differentiable Baryon Pasting framework offers a route to jointly constraining feedback efficiency, non-thermal pressure support, and cosmological parameters, directly improving the reliability of cluster-based cosmological inference.

\begin{acknowledgments}

We thank Kritti Sharma, Ryuichi Takahashi, Masato Shirasaki, and Masaya Yamamoto for comments on the manuscript and for generously sharing their observational data, and Liam Connor for stimulating discussions. We thank Erwin Lau for technical support with the Baryon Pasting code.
This work is supported by the National Science Foundation (NSF) grant AST 2511137 and the Yale Center for Research Computing facilities and staff. I.M. acknowledges support from the Dean's Emerging Scholars Research Award from the Yale Graduate School of Arts \& Sciences.
\end{acknowledgments}

\bibliography{references}{}
\bibliographystyle{aasjournalv7}

\end{document}